\documentclass{aa}
\usepackage{txfonts}
\usepackage{graphics}
\usepackage{epsfig}

\begin{document}

  \title{The metallicity gradient as a tracer of history and
  structure: the Magellanic Clouds and M33 galaxies}

  \author{Maria-Rosa L. Cioni}

  \offprints{M.Cioni@herts.ac.uk}

  \institute{University of Hertfordshire, Science and Technology
  Research Institute, Hatfield AL10 9AB, United Kingdom}

  \date{Received 24 March 2009 / Accepted 19 August 2009}

  \titlerunning{Metallicity gradient}

  \authorrunning{M.-R.L. Cioni}

  \abstract{The stellar metallicity and its gradient pose constraints
    to the formation and evolution of galaxies.}{This is a study of
    the metallicity gradient of the LMC, SMC and M33 galaxies derived
    from their asymptotic giant branch (AGB) stars.}{The [Fe/H]
    abundance was derived from the ratio between C- and M-type AGB
    stars and its variation analysed as a function of galactocentric
    distance. Galaxy structure parameters were adopted from the
    literature.}{The metallicity of the LMC decreases linearly as
    $-0.047\pm0.003$ dex kpc$^{-1}$ out to $\sim 8$ kpc from the
    centre. In the SMC, [Fe/H] has a constant value of $\sim
   -1.25\pm0.01$ dex up to $\sim 12$ kpc. The gradient of the M33 disc,
    until $\sim 9$ kpc, is $-0.078\pm0.003$ dex kpc$^{-1}$ while an
    outer disc/halo, out to $\sim 25$ kpc, has [Fe/H]$\sim -1.7$
    dex.}{The metallicity of the LMC, as traced by different
      populations, bears the signature of two major star forming
      episodes: the first one constituting a thick disc/halo
      population and the second one a thin disc and bar due to a close
      encounter with the MW and SMC. The [Fe/H] of the recent episode
      supports an LMC origin for the Stream. The metallicity of the
      SMC supports star formation, $\sim 3$ Gyr ago, as triggered by
      LMC interaction and sustained by the bar in the outer region of
      the galaxy. The SMC [Fe/H] agrees with the present-day abundance
      in the Bridge and shows no significant gradient. The metallicity
      of M33 supports an ``inside-out'' disc formation via accretion
      of metal poor gas from the interstellar medium.}

\maketitle

\section{Introduction}
\label{intro}
The Magellanic Clouds and M33 galaxies are members of the Local
Group. They contain many AGB stars that have been used to study their
star formation history (SFH) and structure (e.g. Cioni, Habing \&
Israel \cite{mor}, van der Marel \& Cioni \cite{vdm}, Cioni et
al \cite{cmr}, \cite{lfl}, \cite{lfs}, \cite{m33}).  AGB stars exist
in two forms: carbon-rich (C-type) and oxygen-rich (M-type), depending
on the chemical abundance (atoms and molecules) of their
atmosphere. Their ratio, the C/M ratio, is an established indicator of
metallicity and the most comprehensive, whilst not perfect,
calibration as a function of [Fe/H] is given by Battinelli and Demers
(\cite{bat}).

The investigation of the metallicity gradient in galaxies is directly
linked to their formation mechanism. Generally, in a collapse scenario
gas is accreted and falls into the centre where stars form, enriching
the pre-existing gas. Stars may also form during the accretion process
at a given distance from the centre. Bar, disc and halo components as
well as the dynamical interaction of galaxies and the accretion of
satellites alter the distribution of gas. The detection of
metallicity, and age, gradients is crucial to interpret the formation
and evolution mechanisms.

Gradients in total metallicity, iron, oxygen or $\alpha$ elements may
differ because these do not share the same origin and evolution, and
therefore trace different moments in the history of
galaxies. According to stellar evolution theory, iron-peak elements
are mostly produced in the explosion of supernova (SN) of type I, with
low- and intermediate-mass stars progenitors. Oxygen and other
$\alpha$ elements are instead primarily produced by SN type II, with
massive stars progenitors.  The relation between iron and oxygen
depends on the galaxy.

The Magellanic Clouds are a pair of interacting galaxies.  Recent
measurements of their proper motion (Kallivayalil et al \cite{kal06a},
\cite{kal06b}) suggest that they are approaching the Milky Way (MW)
for the first time (Besla et al \cite{bes07}). Their mutual
interaction, rather than the interaction with the MW, is fundamental
in shaping their SFH and metallicity gradients. The Large Magellanic
Cloud (LMC) is a late-type spiral galaxy seen nearly face-on, rich in
gas and with active star formation while the Small Magellanic Cloud
(SMC) is a highly inclined irregular galaxy with less active
star formation. Their dynamical interaction is claimed responsible of
the various star forming episodes and of the creation of the
Magellanic Bridge, connecting the two galaxies, (Gordon et al
\cite{gor09}) and Stream (Nidever et al \cite{nid08}).  The LMC is
probably just a few kpc thick, along the line of sight, but the SMC
has a more complex structure that may extend up to $20$ kpc. Their
apparent morphology is dominated by the distribution of young stars
while evolved stars trace a more regular elliptical structure (Cioni,
Habing \& Israel \cite{mor}). Embedded in each galaxy is a bar. The
Magellanic Clouds have experienced an extended SFH (e.g.~Hill
\cite{hil00}, Zaritsky et al \cite{zar02}, \cite{zar04}; Cole et al
\cite{col05}, Pomp\'{e}ia et al \cite{pom08}, Gallart et al
\cite{gal08}, Carrera et al \cite{car}, \cite{car08a}).

M33 is an isolated spiral galaxy. Its most prominent feature is a
warped disc embedding well delineated spiral arms. Surface brightness
profiles indicate that the disc is truncated at $\sim 8$ kpc (Ferguson
et al \cite{fer07}) and while there might be a halo (Schommer et al
\cite{scho91}, McConnachie et al \cite{mc06a}, Sarajedini et al
\cite{sar06}) there is no bulge (McLean \& Liu \cite{mcl96}).  The SFH
of the inner disc is different from that of the outer disc/halo
(e.g. Barker et al \cite{bar07}, Williams et al
\cite{wil09}). Inhomogeneities in age and metallicity have been
presented by Cioni et al (\cite{m33}).

This paper, motivated by previous investigations, explores the [Fe/H]
abundance variation with galactocentric distance in the Magellanic
Clouds and M33. Section \ref{data} describes the AGB samples, the
calculation of [Fe/H] and of distances as well as of
gradients. Section \ref{dis} discusses these gradients with respect to
the literature and the implication on the structure, formation and
evolution of each galaxy.  Section \ref{concl} concludes this
study. The appendix discusses the iron abundance with respect to the
Ca II triplet (Sect.~\ref{app}) and the C/M ratio (Sect.~\ref{cmfe}).

\section{Analysis}
\label{data}

\subsection{The AGB sample}
\label{sample}
The data analysed here are from Cioni \& Habing (\cite{cmr}), for the
Magellanic Clouds, and Cioni et al (\cite{m33}), for M33.  The samples
comprise $32801$ and $7653$ AGB stars within two areas of $20\times20$
deg$^2$ in the LMC and SMC, respectively, and $14360$ in M33 within
$1.8\times1.8$ deg$^2$.  The LMC and SMC areas, centred at
$(\alpha=82.25^\circ,\delta=-69.5^\circ)$ and
$(\alpha=12.5^\circ,\delta=-73^\circ)$, were divided using a grid of
$100\times100$ cells of size $0.04$ deg$^2$ each. The M33 area,
centred at $(\alpha=23.46^\circ,\delta=30.66^\circ)$, was divided
using a grid of $36\times36$ cells of size $0.0025$ deg$^2$ each.

The number of C- and M-type AGB stars was selected using
colour-magnitude diagrams. DENIS-$IJK_s$ data were used for the
Magellanic Clouds. This combination of optical and near-infrared
broad-band observations allows to minimize the contamination of the
AGB sample by foreground and red giant branch (RGB) stars. The reader
should refer to Cioni \& Habing (\cite{cmr}) for the sub-type of AGB
stars selected. Only $JK_s$ data were available for M33 and the most
reliable, whilst not complete, sample of AGB stars was obtained by
selecting AGB stars above the tip of the RGB that were classified as C
if $(J-K_s)_0>1.36$ and as M if comprised within two slanted lines,
following the shape of the giant branch and in agreement with
theoretical stellar evolutionary models (see Cioni et al \cite{m33}).

Prior to the selection of the AGB sample the data were dereddend
  to account for interstellar extinction along the line of sight, as
  in Cioni et al (\cite{lfl}, \cite{lfs} and \cite{m33}). If
  differential reddening is, however, present throughout a galaxy this
  may affect the source selection based on colours and magnitudes. 
The extinction map derived by Zaritsky et al (\cite{zar04}) for the
LMC shows that dust is highly localized near young and hot stars while
there is no global pattern. The extinction towards older stars is
bimodal reflecting the location of stars in front and behind a thin
dust layer embedding the young stars. This extinction corresponds to
absorption peaks of $A_J=0.02$, $0.12$ and $A_K=0.01$, $0.05$ for
$A_V=0.1$, $0.5$ and adopting the Glass \& Schultheis (\cite{gla03})
extinction law.  The lack of pattern does not influence the global
shape of the galactocentric trends discussed here, but it will
introduce scatter around them. No extinction map is at present
available for the SMC and M33.

\subsection{The iron abundance}
\label{iron}
The C/M ratio has been established as a good indicator of metallicity.
Battinelli \& Demers (\cite{bat}) have provided a relation to convert
this ratio into iron abundance: [Fe/H]$=-1.32\pm0.07 -0.59\pm0.09
\times log(C/M0+)$. This relation was obtained by classifying
homogeneously AGB stars in a wide sample of Local Group galaxies and
by adopting [Fe/H] values from, mostly, RGB stars. The latter
represent the closest approximation to the metallicity of the AGB
progenitors across their age range.  A re-assessment of this relation,
[Fe/H]$=-1.39\pm0.06 -0.47\pm0.10 \times log(C/M0+)$, is given in
Sect.~\ref{cmfe}.

\begin{figure}
\resizebox{\hsize}{!}{\includegraphics{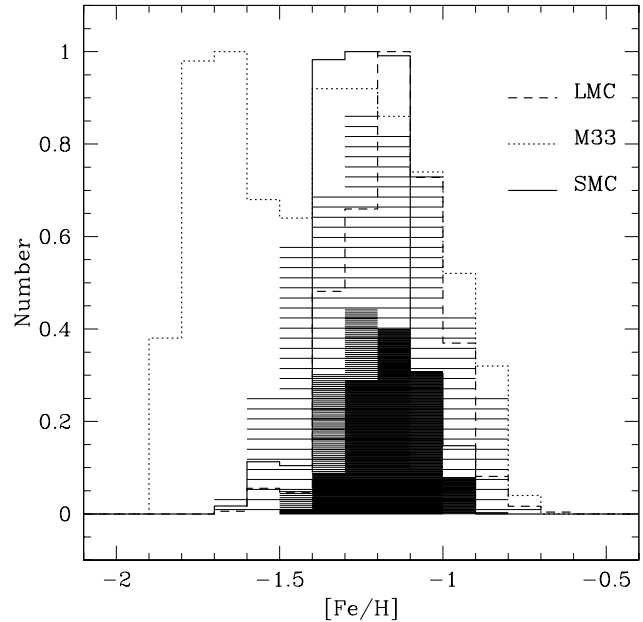}}
\caption{Metallicity distribution for the LMC (dashed line), M33
  (dotted line) and the SMC (continuous line) normalized to their peak
  of $544$, $115$ and $50$ stars, respectively. Shaded areas indicate
  [Fe/H] values with $\sigma_\mathrm{[Fe/H]}<0.2$ dex, different
  patterns correspond to the LMC (filled lines), SMC (narrow-spaced
  lines) and M33 (wide-spaced lines).}
\label{histo}
\end{figure}

The C/M ratio obtained within each cell across the Magellanic Clouds
and M33 has been converted into [Fe/H] using this relation. The
uncertainty in the resulting values of [Fe/H] is the propagated error
on the parameters that characterize the relation and on the error on
the number of C and M stars, the square root of these numbers. This
error is $\ge 0.1$ if C/M$\ge 0.2$ or [Fe/H]$\le1.4$ dex. Figure
\ref{histo} shows the metallicity distribution across each galaxy. The
distribution of M33 is bimodal with peaks at [Fe/H]$\approx -1.3$ and
[Fe/H]$\approx -1.65$ dex. The LMC peaks at [Fe/H]$\approx -1.15$ dex
and the SMC at [Fe/H]$\approx -1.25$ dex. Sources with
$\sigma_\mathrm{[Fe/H]}<0.2$ dex populate mostly the metal-rich peak
in M33, the metal-poor peak appears, however, significant with respect
to the uncertainties involved.  The metallicity of the LMC is higher
than that of the SMC while the metal-rich M33 peak is wide and
encompasses both Magellanic peaks.

\subsection{The AGB gradient}
\label{grad}

\begin{table}
\caption{Parameters of galaxy structures}
\label{val}
\[
\begin{array}{ccccc}
\hline
\noalign{\smallskip}
\mathrm{Galaxy} & \mathrm {Distance} & \mathrm{Pos.~Angle}
& \mathrm{Inclination} & \mathrm{Reference(s)} \\
 & \mathrm{(kpc)} & \mathrm{(deg)} & \mathrm{(deg)} & \\
\noalign{\smallskip}
\mathrm{LMC} & 51 & 189.3 & 34.7 & 1,5,3\\
\mathrm{SMC} & 63 & 45 & 65.5 & 1,4\\
\mathrm{M33} & 840 & 22 & 53 & 2\\
\noalign{\smallskip}
\hline
\end{array}
\]
($1$) Cioni et al (\cite{tip}); ($2$) Magrini et al (\cite{mag09});
($3$) van der Marel \& Cioni (\cite{vdm}); ($4$) Paturel et al
(\cite{pat}); ($5$) van der Marel (\cite{vdm1}).
\end{table}

A galactocentric distance, $R_{GC}$, has been associated to each
cell. These values were derived using: distance ($D$), position angle
of the major axis ($PA$) and inclination ($i$), as listed in
Tab.~\ref{val} and proceeding as follows:
\vspace{-0.2cm}
\begin{description}
\item -- convert the equatorial coordinates of each star
  ($\alpha_i,\delta_i$) into angular coordinates ($x_i,y_i$) 

\item -- group stars into cells of a given grid (Sect.~\ref{sample})

 \item -- rotate the coordinate system according to:
\vspace{-0.2cm}
\begin{equation}
x1 = x \times cos(\theta)+y \times sin(\theta)
\end{equation}
\vspace{-0.5cm}
\begin{equation}
y1 = y \times cos(\theta)-x \times sin(\theta)
\end{equation}
\vspace{-0.2cm} where ($x,y$) are mid-cell values in deg and
$\theta=\mathrm{PA}-90^{\circ}$

\item -- de-project using:
\vspace{-0.2cm}
\begin{equation}
y2 = y1 / cos(i)
\end{equation}

\item -- calculate the angular distance and convert into kpc with:
\vspace{-0.2cm}
\begin{equation}
d_\mathrm{deg} = \sqrt{x1^2+y2^2}
\end{equation}
\vspace{-0.5cm}
\begin{equation}
d_\mathrm{kpc} = D \times tg (d)
\end{equation}
\vspace{-0.2cm}
where $d$ is the angular distance of each cell.
\end{description}

No uncertainty was calculated for the resulting $R_\mathrm{GC}$ values
due to the unknown uncertainty on the parameters that describe the
structure of the galaxies. It is likely that the uncertainty on
$R_\mathrm{GC}$ will be dominated by the uncertainty on the distance
to the galaxy, especially where the thickness along the line of sight
is not known, e.g. SMC.  The distribution of [Fe/H] is shown in
Fig.~\ref{lmcgrad}, \ref{smcgrad} and \ref{m33grad} for the LMC, SMC
and M33, respectively.

Figure \ref{lmcgrad} shows that the LMC has a smooth gradient where
the central region is more metal rich, [Fe/H]$\sim-1.0$ dex, compared
to the outer region, [Fe/H]$\sim -1.5$ dex at $\sim 10$ kpc. A
non-weighted least square fit through all points gives
[Fe/H]$=-0.035\pm0.002 \times R_\mathrm{GC} - 1.03\pm0.01$ with a
typical uncertainty on a single measurement of $0.14$ (rms) while if
only points with $\sigma_\mathrm{[Fe/H]}<0.2$ dex are considered the
resulting fit is [Fe/H]$=-0.047\pm0.003 \times R_\mathrm{GC} -
1.04\pm0.01$ with rms$= 0.09$.

A negligible gradient is derived for the SMC (Fig.~\ref{smcgrad}). A
non-weighted least square fit through all points gives
[Fe/H]$=-0.007\pm0.003 \times R_\mathrm{GC} - 1.22\pm0.01$ with rms$=
0.13$ while for $\sigma_\mathrm{[Fe/H]}<0.2$ dex the resulting fit is
[Fe/H]$=0.004\pm0.005 \times R_\mathrm{GC} - 1.25\pm0.01$ with rms$=
0.09$. Note that the slope is different between the two fits but in
both cases it is consistent with a flat gradient.

The M33 gradient has a dual distribution (Fig.~\ref{m33grad}) and the
change in slope occurs at about the truncation radius, $\sim 8$ kpc
(Ferguson et al \cite{fer07}). A non-weighted fit of all points with
$\sigma_\mathrm{[Fe/H]}<0.2$ dex, that are approximately confined
within this radius and have [Fe/H]$\ge -1.5$, gives [Fe/H]$=-0.078\pm
0.003 \times R_\mathrm{GC} - 0.77\pm0.02$ with rms$= 0.11$. Points
with [Fe/H]$<-1.5$ dex have $R_\mathrm{GC}>8$ kpc and their fit is
[Fe/H]$=-0.007\pm0.002\times R_\mathrm{GC}-1.60\pm0.03$ with
rms$=0.09$. By neglecting the dual distribution and the uncertainty
one obtains [Fe/H]$=-0.057\pm0.002 \times R_\mathrm{GC} - 0.93\pm0.02$
with rms$= 0.17$.

\section{Discussion}
\label{dis}

\subsection{The Large Magellanic Cloud}

\begin{figure*}
\resizebox{\hsize}{!}{\includegraphics{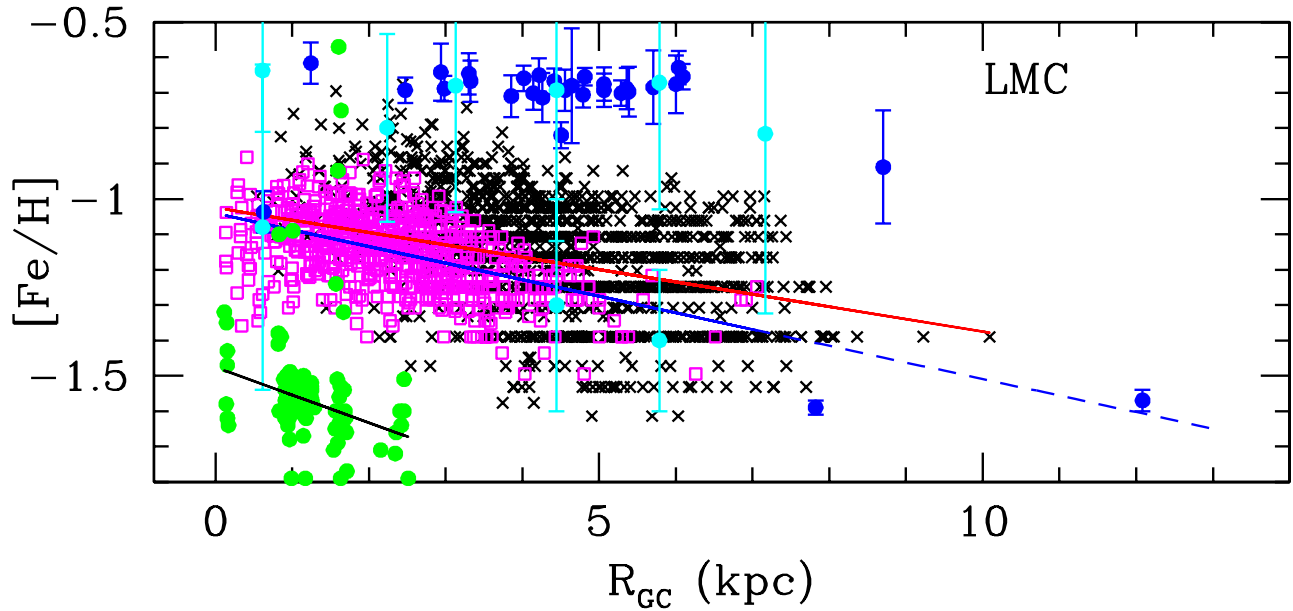}}
\caption{Iron abundance in the LMC. Points referring to AGB abundances
  with $\sigma_\mathrm{[Fe/H]}<0.2$ dex are plotted as empty squares
  (magenta) and those with $0.2\le \sigma_\mathrm{[Fe/H]}<0.38$ as
  crosses (black). The least square fit lines through all data points
  (red) and through only those with small uncertainties (blue) are
  indicated, the latter has been prolongued with a dashed line to $13$
  kpc. Filled circles show RGB abundances by Cole et al
  (\cite{col05}), Pomp\'{e}ia et al (\cite{pom08}) and Carrera et al
  (\cite{car08a}) in the field (light blue) as from Tab.~\ref{met},
  Grocholski et al (\cite{gro06}, \cite{gro07}) for stellar clusters
  (with small error bars; dark blue) and Borissova et al
  (\cite{bor06}) for RR Lyrae stars (without error bars; green), see
  text for details. The colour figure is available electronically.}
\label{lmcgrad}
\end{figure*}
 
Literature studies of the LMC refer to an inner and an outer disc
component simply differentiating how far the observed regions are from
the centre. The presence of an inner/outer halo component is instead
drawn from metallicity measurements. A halo containing predominantly
gas from the initial process of galaxy formation would be metal poorer
than the disc of the galaxy where stars have formed. A metal rich halo
would instead bear the signature of significant accretion of small
bodies. The bar, residing in the disc (Zaritsky et al \cite{zar94}),
is usually referred to as a separate component and can considerably
reduce pre-existing abundance gradients over a few dynamical
timescales since its formation (Friedli \& Benz \cite{fri95}).

\subsubsection{AGB and RGB gradients}

Cole et al (\cite{col05}) derived [Fe/H] using the Ca II triplet
method in a sample of RGB stars in the LMC bar. Carrera et al
(\cite{car08a}) have used the same method for RGB stars at $3-7$ kpc
and Pomp\'{e}ia et al (\cite{pom08}) for RGB stars at $\sim 2$ kpc.
The original data-point from Pomp\'{e}ia et al (\cite{pom08}) is the
mean and standard deviation, [Fe/H]$=-0.75\pm0.23$, of their sample
(Sect.~\ref{app}). The two Cole et al (\cite{col05}) points correspond
one to the disc (metal rich) and the other to the halo (metal poor).
Carrera et al (\cite{car08a}) quote metallicities only for the disc.
On the other hand, their Fig.~5 shows that a halo component exists in
at least two of their fields. I derived the intensity and width of
this component from their histograms. Table \ref{met} shows the
values of [Fe/H] obtained from Cole et al (\cite{col05}), Pomp\'{e}ia
et al (\cite{pom08}) and Carrera et al (\cite{car08a}) original data
and the values resulting from applying a correction for the difference
between Ca II triplet abundances and abundances obtained directly from
iron lines (Sect.~\ref{app}). The latter are used in this study. 

\begin{table}
\caption{[Fe/H] abundances in the LMC}
\label{met}
\[
\begin{array}{cccc}
\hline
\noalign{\smallskip}
\mathrm{Distance} & \mathrm{[Fe/H]}_\mathrm{orig} & 
\mathrm{[Fe/H]}_\mathrm{cor} & \mathrm{Reference} \\ 
\mathrm{(kpc)} & \mathrm{(dex)} & \mathrm{(dex)} & \\
\noalign{\smallskip}
0.6 & -0.37\pm0.15 & -0.64\pm0.17 & 1 \\
0.6 & -1.08\pm0.46 & & 1 \\
2.4 & -0.75\pm0.23 & -0.79\pm0.27 & 2 \\
3.1 & -0.47\pm0.31 & -0.68\pm0.36 & 3 \\
4.4 & -0.50\pm0.37 & -0.69\pm0.43 & 3 \\ 
4.4 & -1.30\pm0.30 & & 3 \\
5.8 & -0.45\pm0.31 & -0.67\pm0.36 & 3 \\ 
4.4 & -1.30\pm0.30 & & 3 \\
7.2 & -0.79\pm0.44 & -0.82\pm0.50 & 3 \\

\noalign{\smallskip}
\hline
\end{array}
\]
($1$) Cole et al (\cite{col05}); ($2$) Pomp\'{e}ia et al
(\cite{pom08}); ($3$) Carrera et al (\cite{car08a}).
\end{table}

Figure \ref{lmcgrad} shows that RGB values, compared to AGB ones, have
a dual behaviour: those of the disc have high abundances and a
negligible gradient out to $\sim 6$ kpc, and those of the halo are
metal poor and follow more closely the AGB gradient. 
The intensity of star forming episodes, the dynamical effect of the
bar and the re-distribution of stars from their birth place
(Ro\v{s}kar et al \cite{ros08}) may be responsible for the different
gradients and for the scatter around them.  It is unlikely that the
photometric criteria used to discriminate between C- and M-type favour
low metallicities. In fact it is somewhat easier to distinguish C
stars from their near-infrared colours and magnitudes because at their
location there are very few sources of contamination (Cioni et al
\cite{cio01}, Battinelli \& Demers \cite{bat09}). The M stars region,
however, relies strongly on the minimization of the contribution by
RGB and galactic dwarf stars. The result is a bias in isolating
preferentially metal-rich regions, with low C/M ratios, as M stars
would be over-estimated compared to C stars within the same region.

Most AGB stars are long period variables (LPVs) and Hughes et al
(\cite{hug91}) derived that $40$\% of them are old ($\ge 9$ Gyr) and
part of a spheroidal population, contrary to $53$\% being of
intermediate age ($\sim 4$ Gyr) and residing in a disc; the others are
young ($\le 1$ Gyr). Most C stars reside on a thick disc (van der
Marel et al \cite{vdm02}) and are $1-4$ Gyr old (Marigo et al
\cite{ma99}).  The distinction between thin and thick disc is not
clear. In our Galaxy thick disc stars have higher $[\alpha/\mathrm
  {Fe}]$ than thin disc stars (Soubiran \& Girard \cite{sou05}) but
$[\alpha/\mathrm {Fe}]$ might follow a gradient within either discs
(Edvardsson et al \cite{edv93}). The low $[\alpha / \mathrm{Fe}]$
ratio measured by Pomp\'{e}ia et al (\cite{pom08}) suggests a higher
contribution by SN type Ia relative to type II supporting the
formation of their stars at intermediate ages. Cole et al
(\cite{col00}) indicate that these RGB stars are likely $1-3$ Gyr old,
such as most of the RGB stars studied by Cole et al (\cite{col05}) and
Carrera et al (\cite{car08a}).  If many of the AGB stars analysed here
are older than the bulk of RGB stars they will more closely follow the
outer disc/halo. 


\subsubsection{Does the LMC have a stellar halo?} 

A break in the surface brightness profile, usually associated with the
transition between two components, is present at $\sim 4^\circ$ ($\sim
3.6$ kpc) van der Marel (\cite{vdm1}). The AGB gradient at
$5<R_\mathrm{GC}<8$ is only marginally flatter than in the inner disc.
Observations of stellar clusters exhibit a disc-like kinematics that
is very similar to the HI disc (Grocholski et al \cite{gro06},
\cite{gro07}).  Their corrected metallicity, using Eq.~\ref{cor}, is
constant at [Fe/H]$=-0.66\pm0.02$ dex, in agreement with the inner
disc metallicity in field RGBs. Grocholski et al (\cite{gro06})
attributed this flattening to the dynamical effect of a bar that
occupies a significant fraction of the disc length. The older and
metal poorer clusters, however, follow the AGB gradient, regardless of
their location with respect to the galaxy centre, this is also true
for RGB stars associated with the halo. The small number of these
clusters does not allow to characterize their kinematics but suggests
that a halo extends at least to $\sim 14$ kpc from the centre, LMC
stars have been claimed at $\sim 20$ kpc (Majewski et al
\cite{maj05}), while studies of AGB stars are limited to $\sim 10$
kpc.

RR Lyrae stars are often attributed to the halo of galaxies because of
their old age and velocity dispersion (Minniti et al \cite{min03}).
In the LMC they may have formed in the disc and subsequently moved to
the halo as a consequence of a merger event in the early formation of
the galaxy (Subramaniam \cite{sub06}). This halo formed before the
disc currently traced by red clump giant stars (Subramanian \&
Subramaniam \cite{sub09}).  The metallicity of RR Lyrae stars measured
by Borissova et al (\cite{bor06}) and their least square fit,
[Fe/H]$=-0.078\pm0.007\times R_\mathrm{GC}-1.48\pm0.03$ with
rms$=0.18$, are shown in Fig.~\ref{lmcgrad}. This gradient is steeper
than that obtained from AGB stars and population II clusters ($\ge 9$
Gyr old).  The latter is steeper than that from disc RGB stars and
intermediate-age ($1-3$ Gyr) clusters.

\subsubsection{Chemical enrichment and dynamics}

The difference between the AGB and the RR Lyrae stars gradients,
supported by old clusters, suggests that chemical enrichment has
occured between the formation of their progenitors. If these AGB stars
are old (Hughes et al \cite{hug91}) a $2$ Gyr difference would produce
an enrichment of $\sim 0.02$ dex kpc$^{-1}$ Gyr$^{-1}$. On the other
hand, the mean age difference between RR Lyrae stars and young RGB
stars or young clusters corresponds to $8$ Gyr. By comparing their
gradients we obtain an enrichment of $\sim 0.01$ dex kpc$^{-1}$
Gyr$^{-1}$. If the enrichment took place in the last $3$ Gyr then the
rate can be as high as $0.05$ dex kpc$^{-1}$ Gyr$^{-1}$. A steep
  enrichment is also supported by the difference between the AGB and
  young-RGB gradients. The metallicity offset between RR Lyrae stars
  and AGB stars is comparable to the difference between AGB stars and
  young RGB stars. This effect can be explained in terms of age
  differences but could also be influenced by the dispersion in the
  calibration of the C/M-[Fe/H] relation as well as uncertainties in
  metallicities obtained via other methods. There are insufficient
data on HII regions and PNe to investigate further the chemical
enrichment of the LMC. The census of these objects is highly
incomplete and biased to younger members (Reid and Parker \cite{pne},
Leisy \& Dennefeld \cite{lei06}, Dufour et al \cite{duf84}). Metal
poor PNe distributed over a halo fall below the sensitivity of
previous studies. High PNe abundances, that do not include iron, are
confined to the bar and south eastern region of the LMC indicating
places were star formation was recently active (Leisy \& Dennefeld
\cite{lei06}).

According to Bekki et al (\cite{bek04}) the LMC experienced a close
encounter with the SMC $\sim 4$ Gyr ago. This event caused a new
episode of star formation in both the field and cluster population as
well as the formation of the LMC bar and Magellanic Stream. The
average metallicity of the Stream is [Fe/H]$=-0.6\pm0.2$ dex (Wakker
\cite{wak01}) suggesting a similar age as that of disc RGB and cluster
stars (Fig.~\ref{lmcgrad}). Nidever et al (\cite{nid08}) argue for an
LMC origin of the Stream that is $\sim 2$ Gyr old. This is consistent
with ram-pressure gas stripping from the outer LMC disc due to a close
passage by the MW (Mastropietro \cite{mas08}). The new LMC orbit,
derived from the new proper motion (Kallivayalil et al \cite{kal06a}),
implies that the LMC passed perigalacticon $\sim 1.78$ Gyr ago. The
AGB gradient analysed here bears the signature of the star forming
episode, several Gyr ago, responsible for the formation of a thick
disc/halo component. The metal-rich RGB and the cluster stars are the
product of this recent episode while RR Lyrae stars, some RGB stars
and clusters refer to the earliest episode.

\subsection{The Small Magellanic Cloud}

\begin{figure*}
\resizebox{\hsize}{!}{\includegraphics{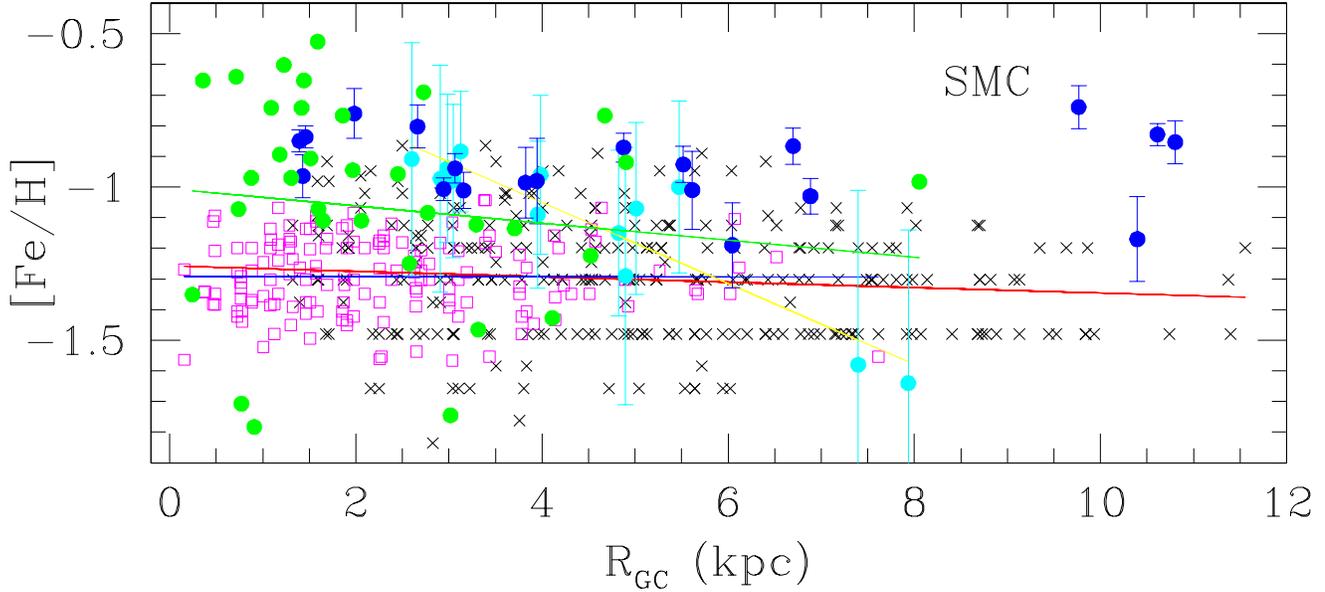}}
\caption{Iron abundance in the SMC. Points are as in
  Fig.~\ref{lmcgrad}. The least square fit lines through all points
  (red) and only those with small uncertainties (blue) are
  indicated. Filled circles with large error bars (light blue) and
  their least square fit line (yellow) refer to RGB stars by Carrera
  et al (\cite{car}) while filled circles with small error bars (dark
  blue) are for stellar clusters (Parisi et al \cite{par09}, Da Costa
  \& Hatzidimitriou \cite{dac98}). Both measurements were corrected as
  explained in the text. Filled circles without error bars and the
  best fit line through them (green) refer to PNe by Idiart et al
  (\cite{idi07}). The colour figure is available electronically.}
\label{smcgrad}
\end{figure*}

\subsection{Flat gradient or metal-rich ring?}

 In Cioni et al (\cite{lfs}) it was recognized that the SMC bar region
 is surrounded by a metal-rich ring with signatures of dynamical
 evolution, moving clumps, as a function of time. This behaviour was
 un-explained and attributed to the unknown geometry of the galaxy. A
 ring feature occurring at $\sim 2.5$ Gyr and persisting until $\sim
 1.6$ Gyr was found by Harris \& Zaritsky (\cite{har04}) in their SFH
 analysis. Its age agrees with the age of AGB stars, $0.6-2$ Gyr old
 (Cioni et al \cite{smcvar}).  Before and after, star formation
 occurred in the bar. In both studies the metallicity was derived from
 stellar evolution models in terms of Z that represents the total
 heavy element abundance.  A flat [Fe/H] gradient would be, then,
 consistent with an $\alpha$- or O-rich ring.


Very recently, Gonidakis et al (\cite{god09}) suggested that the old
(K, M and faint C stars) stellar population rotates and resides on an
a disc. Combining this information with a central region that started
to form stars $\sim 10$ Gyr ago it is possible that a metal rich ring
is the result of star formation induced by a rotating bar, that
sustains gas in the outer parts of the galaxy. The rotation speed of
the bar may be responsible for an age gradient in the ring. The
presence of a bar is also consistent with a flat metallicity gradient
because it acts against a linearly decreasing gradient that would
instead be present in a bar-less disc galaxy (Martin \& Roy
\cite{roy94}). A bar, regardless of the overall structure of the
galaxy (spiral, spheroidal) always resides in a disc (Zaritsky et al
\cite{zar94}). On the other hand, the investigation by Subramaniam \&
Subramanian (\cite{sub09}) shows that the SMC might host a bulge where
metal poor and metal rich stars coexist and trace a similar line of
sight depth. Although the two concepts do not exclude each other
(e.g. Galactic bulge) more evidence is needed to confirm these
sub-structures.

\subsubsection{Halo and chemical enrichment}

The data analysed here do not show evidence for a halo population
traced by AGB stars. In the literature there is no evidence for such a
halo, although there have been a limited number of studies suggesting
that the extent of the SMC disc is larger than the, more familiar,
optical appearance of the galaxy (e.g. N\"{o}el \& Gallart
\cite{noe07}). There is also an apparent lack of metallicity
determinations in RR Lyrae stars, usually populating halos. 

The [Fe/H] distribution of stellar clusters (Parisi et al
\cite{par09}, Da Costa \& Hatzidimitriou \cite{dac98}) shows also a
flat gradient.  The cluster metallicities plotted in
Fig.~\ref{smcgrad} are those obtained after applying Eq.~\ref{cor} to
the original data according to App.~\ref{app}. The correction might
not be appropriate for the Da Costa \& Hatzidimitriou (\cite{dac98})
values, that rely on a different calibration, but their agreement with
the Parisi et al (\cite{par09}) measurements is independent on this
correction. Most clusters are of intermediate age, $2.7\pm1.6$
Gyr. There are only three old clusters, $\ge 9$ Gyr, similar to those
found in the metal poor outer disc/halo of the LMC but their
metallicity does not differ from that of younger clusters.

 RGB metallicities by Carrera et al (\cite{car}) suggest a steep
gradient, [Fe/H]$=-0.13\pm0.02 \times R_\mathrm{GC} - 0.52\pm0.08$
with rms$= 0.10$ (Fig.~\ref{smcgrad}), after applying Eq.~\ref{cor} to
those fields dominated by a stellar population younger than $5$ Gyr.
The two fields that drive the steepness of the gradient might not be
representative of all position angles at a similar de-projected radii
in the SMC and have very large error bars.  A fit that excludes these
points is shallower, [Fe/H]$=-0.08\pm0.03 \times R_\mathrm{GC} -
0.71\pm0.11$ with rms$= 0.08$. The difference betwen the AGB gradient
and the gradient from young RGB stars implies a chemical enrichment of
$0.04$ dex kpc$^{-1}$ Gyr$^{-1}$ for an average age difference of $2$ Gyr. 


Idiart et al (\cite{idi07}) measured [O/H] in many PNe and converted
it into [Fe/H] using: [Fe/H]$=0.16+1.27\times$[O/H]. By de-projecting
the coordinates of each PNe as in this study a gradient of
$-0.03\pm0.05$ dex is obtained (Fig.~\ref{smcgrad}). These PNe are
mostly located within $5$ kpc except one at $8$ kpc. It is interesting
to note that with the solar abundance calibration by Asplund et al
(\cite{asp04}) PNe are metal-richer than AGB stars by $\sim 0.3$ dex
while with the Anders \& Grevesse (\cite{and89}) calibration there is
good agreement. In the calculation of the PNe gradient I excluded
objects with depressed oxygen abundance and four others for which I
could not find coordinates. Figure \ref{smcgrad} shows $39$ PNe, with
no distinction between type I and II, and their metallicities are
consistent with a flat gradient, but it is somewhat steeper than that
obtained from AGB stars. AGB stars are the precursors of PNe and an
age difference between the two is on average negligible. Known PNe
are, however, biased towards the largest, most asymmetric and luminous
members, that have intermediate-mass stars progenitors (Jacoby \& De
Marco \cite{jac02}).  In addition, the initial metallicity derived
from the oxygen abundance of PNe is questionable (Leisy \& Dennefeld
\cite{lei06}) and it would be more appropriate to use elements that
are not modified during the AGB evolution.  In the SMC only a small
number of HII regions have been observed with the aim of determining
their chemical abundance (Dufour \cite{duf84}).  Their [O/H] abundance
agrees with that measured in PNe and shows a small scatter, $0.08$
dex. A larger sample of both HII regions and PNe, is needed before
relating them to the chemical enrichment of the galaxy. Reid \& Parker
(\cite{pne}) have shown, for the LMC, that many PNe await to be
discovered.

\subsubsection{Dynamics and the Magellanic Bridge}

The age-metallicity relation as measured in different SMC fields
(Carrera et al \cite{car}) suggests that after an initial episode of
gas enrichment ($\le 7$ Gyr ago; Tosi et al \cite{tos08}) the SMC
experienced a period of quiescent evolution before a new episode $\sim
3$ Gyr ago. At that time, the SMC had a close encounter with the LMC
(Piatti et al \cite{pia05}, Bekki et al \cite{bek04}). Was this
episode responsible for the formation of the Magellanic Bridge?  The
present-day metallicity of the Bridge, from early-type stars, is
[Z/H]$=-1.05\pm0.06$ dex (Lehner et al \cite{leh08}), where Z includes
oxygen, nitrogen and argon abundances. This value is in good
  agreement with the mean metallicity shown in Fig.~\ref{smcgrad},
  supporting an SMC origin for the Bridge dating back a few Gyrs. The
  [Fe/H] found for the LMC AGBs is similar to the SMC ones and it is
  not excluded that the Bridge has been also influenced by LMC
  material at an earlier epoch. 


Star formation in the Bridge started $100-200$ Myr ago (Harris
\cite{har07}), when the LMC and the SMC had another close encounter
(Kallivayalil et al \cite{kal06a}). This is the same time at which the
SMC tail, a tail of tidal origin located within the Bridge, was
stripped from the galaxy (Gordon et al \cite{gor09}). This means that
if the Bridge material were stripped earlier it did not form
stars. This scenario would support the very low metal abundances
derived by Dufton et al (\cite{duf08}), [Fe/H]$-1.73\pm0.15$ dex from
B-type stars. This value is much lower than any of the values in
Fig.~\ref{smcgrad}. Note that the LMC metallicity at $\sim 5$ kpc from
the centre, close to the break radius, corresponds to the metallicity
of the SMC. Was the LMC gas accreted from the SMC via the Bridge? In
the interaction process between the galaxies it is not unreasonable to
expect that the smaller galaxy will suffer more severe disruption in
its outer parts than the largest galaxy. On the other hand, the outer
parts of the largest galaxy will be more loosely bound than its
central parts and prone to disruption. A more systematic investigation
of the stellar population in the Bridge and in the outer region of the
Magellanic Clouds, will provide a clearer view of the history of their
interaction.

\subsection{M33}

\begin{figure*}
\resizebox{\hsize}{!}{\includegraphics{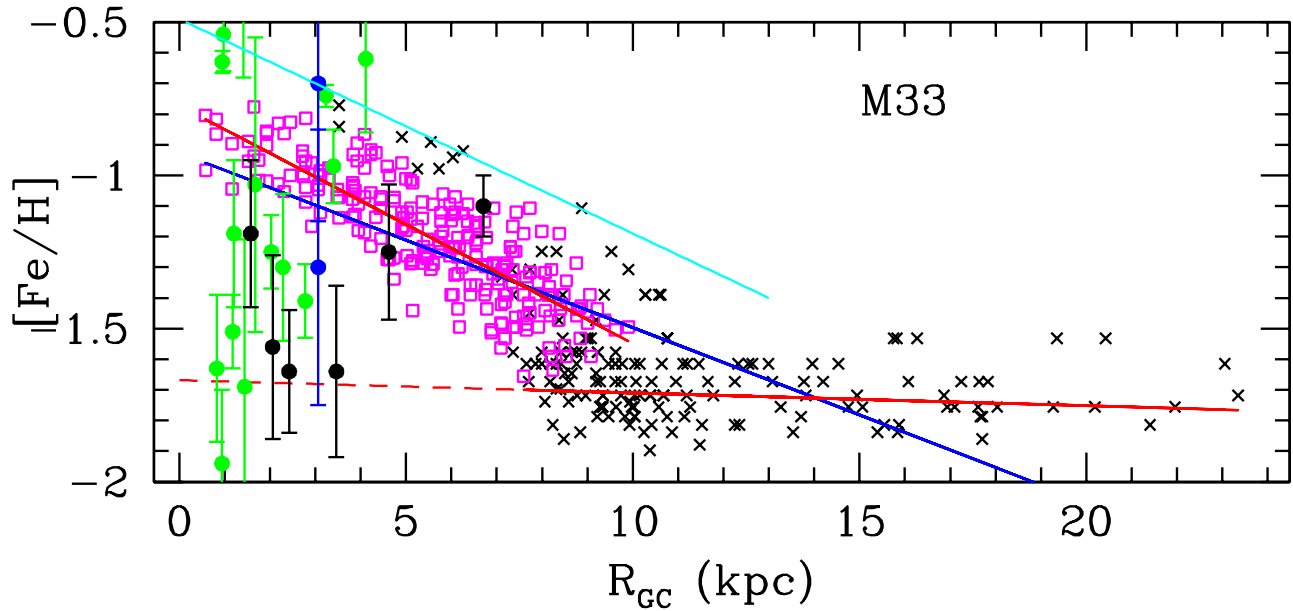}}
\caption{Iron abundance in M33. Points are as in
  Fig.~\ref{lmcgrad}. The least square fit line thorough all points
  (blue) and those with small uncertainties (red) are
  indicated. Magenta points are confined within $\sim 8$ kpc, the
  truncation radius of M33 (Ferguson et al \cite{fer07}). Beyond this
  radius and for [Fe/H]$<-1.5$ dex the least square fit (red) is
  shallower; this fit has been prolonged towards the centre of the
  galaxy with a dashed line.  Filled circles (green \& black) show the
  abundance of stellar clusters attributed to the halo of the galaxy
  (Ma et al \cite{ma} \& Sarajedini et al \cite{sar00}) and to the RR
  Lyrae star population (blue) from Sarajedini et al
  (\cite{sar06}). The RGB gradient derived by Barker et al
  (\cite{bar07}) is shown in cyan. The colour figure is available
  electronically.}
\label{m33grad}
\end{figure*}

The metallicity distribution of M33 is bimodal and the dual gradient
suggests that the galaxy has two clearly distinct components, an inner
disc with a typical linearly decreasing gradient away from the centre,
up to $\sim 9$ kpc, and an outer disc/halo population that dominates
beyond this distance with a much shallower gradient. Barker et al
(\cite{bar07}) tentatively conclude that beyond $50^{\prime}$ ($\sim
13$ kpc) the metallicity gradient flattens but the analysis supporting
this behaviour (Brooks et al \cite{bro04}, Davidge \cite{dav03}) were
highly contaminated from foreground galactic stars and background
galaxies.  Ferguson et al (\cite{fer07}) identified a break in the
surface brightness profile of M33 at $\sim 8$ kpc that nicely
corresponds with the change in the slope of the AGB gradient
suggesting the existence of two components.  The analysis of stellar
clusters (Sarajedini et al \cite{sar00}, Ma et al \cite{ma}) and RR
Lyrae stars (Sarajedini et al \cite{sar06}) support this
scenario. Some clusters de-project onto the inner disc gradient while
others do agree, within the errors, with an outer disc/halo
metallicity.  Furthermore, Chandar et al (\cite{cha02}) by analysing
the kinematics of clusters concluded that old halo candidates have a
[Fe/H] range between $-1.0$ and $-2.0$. This is consistent both with
the plateau derived here beyond $\sim 8$ kpc and with the metal-poor
peak of the metallicity distribution of the Galaxy's globular clusters
(Armandroff \cite{arm99}).

\subsubsection{Chemical enrichment and dynamics}

The RGB gradient corresponds to $-0.07$ dex kpc$^{-1}$ (Kim et al
\cite{kim02}, Tiede et al \cite{tie04}, Barker et al \cite{bar07})
comparable with that derived from AGB stars. If no distinction is made
between inner disc and outer disc/halo AGB stars (Fig.~\ref{m33grad})
a gradient of $-0.057\pm0.002$ is obtained, this is $\sim 0.01$ dex
shallower than the RGB gradient.  Considering that AGB stars span a
large range of ages this indicates that the galaxy did not experience
significant metal enrichment between several to a few Gyrs.
 If AGB stars were younger than the RGB stars they would provide
  evidence for a flattening of the metallicity gradient with time or
  viceversa. The latter might explain the offset between the RGB and
  AGB gradients, but it may also be influenced by the dispersion in
  the C/M-[Fe/H] relation. 

Recent measurements of the gradient from HII regions, Magrini et al
\cite{mag07}, Rosolowsky \& Simon \cite{rosi08}, point to a flat
gradient, [O/H]$\sim -0.03$ dex kpc$^{-1}$, that produces a similar
trend in [Fe/H] using King (\cite{king}) conversion.  HII regions
trace the present-day star formation and the difference between their
gradient and that of the certainly older AGB stars indicates a
flattening of the metallicity gradient with time. Magrini et al
(\cite{mag09}) found that HII regions and PNe follow the same [O/H]
gradient. On the one hand, galaxy chemical evolution models indicate a
steeper gradient for iron than for oxygen simply because iron comes
predominantly from slowly evolving SN type I that compared to SN type
II have not yet enriched the outer parts of galaxies. On the other
hand, the similarity between the HII regions and PNe gradients may
suggest that the PNe sample ($\sim 70$ older than $0.3$ Gyr) is on
average younger than the AGB population ($\sim 14000$) in the disc.

Very recently Williams et al (\cite{wil09}) have shown that the age of
the population in the inner disc decreases radially contrary to the
trend in the outer disc (Barker et al \cite{bar07}, Cioni et al
\cite{m33}). This provides evidence for an inside-out formation
scenario for the M33 disc also supported by simulations. How the
metallicity gradient fits into this picture? Since the majority of the
stars near the centre of the disc had formed by $z=1$ and the bulk of
the stars farther out formed later (Williams et al \cite{wil09}), more
time was available to enrich the gas in the centre than in the outer
parts, heavy elements are also found in the centre because the
potential is stronger. The peak of star formation responsible for
younger ages at about the truncation radius corresponds to $\sim 2$
Gyr; (Williams et al \cite{wil09}; Fig.~3). This suggests that AGB
stars formed, at that time and at that location as a consequence of
accretion of metal poor gas, out of gas that was not enriched in iron
by previous star forming episodes, in agreement with a linearly
decreasing gradient throughout the inner disc region. These AGB stars
are, therefore, older than $2$ Gyr.

The galactic chemical evolution models for the formation of the M33
disc by Magrini et al (\cite{mag07}) indicate a steeper metallicity
gradient in the centre, $\sim -0.11$ dex kpc$^{-1}$, than in the outer
parts of the galaxy, according to a gas accretion model, with an
almost constant gas in-fall rate. The models by Chiappini et al
(\cite{chi01}), for the MW, assume two main accretion episodes: the
first forming the halo/thick disc and the second forming the thin
disc. In their models the disc forms ``inside-out'', in agreement with
the Williams et al (\cite{wil09}) results. A similar picture was
deduced for the formation of the isolated spiral galaxy NGC 300,
member of the Sculptor group and very similar to M33, showing a
negative [Fe/H] gradient in the disc and a flat or slighly positive
gradient in the outer parts (Vlaji\'{c} et al \cite{vla09}). The
alternative explanation populates the outer parts with stellar
migration, accounting for the strength of spiral waves (Ro\v{s}kar et
al \cite{ros08}, Sellwood \& Binney \cite{sel02}).

\section{Conclusions}
\label{concl}

This paper derives the metallicity, [Fe/H], for a large sample AGB
stars in the Magellanic Clouds and M33 and investigates the spatial
gradient with respect to the structure and history of each galaxy as
well as other indicators.  The values for iron abundance depend
strongly on the calibration of the $log(C/M)$-[Fe/H] relation. 
  The metallicity in this relation is that of the dominant population
  of RGB stars that represents the closest approximation, or a lower
  limit, to the metallicity of the AGB progenitors. The relation
provided by Battinelli \& Demers (\cite{bat}) has been revised in this
study to: [Fe/H]$=-1.41\pm0.04 -0.41\pm0.05 \times log(C/M0+)$. The
resulting gradients provide new constraints for theoretical models for
the formation and evolution of these galaxies, and for similar systems
where stars cannot yet be resolved.

The metallicity of the LMC decreases away from the centre,
[Fe/H]$=-0.047\pm0.003 \times R_\mathrm{GC} - 1.04\pm0.01$.  This AGB
gradient is somewhat flatter than that derived from RR Lyrae stars,
$-0.078\pm0.007$ dex kpc$^{-1}$, and it is followed by metal poor
stellar clusters and metal poor RGB stars supporting an old and
extended (up to $14$ kpc) thick disc or halo population. Most RGB
stars and stellar clusters are, however, younger and with a constant
metallicity, [Fe/H]$=-0.66\pm0.02$ dex. They probably formed when the
LMC interacted with the MW and SMC a few Gyr ago along with the
formation of the LMC bar and of the Stream (Nidever et al
\cite{nid08}).  A flattening of the gradient with time is consistent
with ``inside-out'' disc formation (Vlaji\'{c} et al \cite{vla09}),
while a dual formation scenario for the halo and the disc reproduces
the AGB gradient (Chiappini et al \cite{chi01}).


The metallicity of the SMC is consistent with a flat
distribution. This result is sustained by different stellar
indicators: RGB stars, PNe and clusters regardless of their
age. Together with a [M/H]-rich ring (Cioni et al \cite{lfs}; Harris
\& Zaritsky \cite{har04}) and a rotating old stellar population
residing on a disc (Gonidakis et al \cite{god09}) they support the
idea that gas shocked, during an encounter with the LMC $\sim 3$ Gyr
ago, started to form stars in the outer parts of the galaxy altering
the classical [Fe/H] gradient of a bar-less disc galaxy.  Furthermore
an increase in the [$\alpha/$Fe] ratio in the outermost regions of the
disc or a flat [Fe/H] gradient due to equal timescale for disc
formation versus distance are also possible scenarios (Chiappini et al
\cite{chi01}). The [Fe/H] abundance of the SMC agrees with the
present-day abundance in the Bridge.

The M33 inner disc extends to $\sim 9$ kpc, in agreement with previous
findings, while an outer disc/halo population reaches $\sim 25$
kpc. The inner disc is characterised by a steep metallicity gradient,
[Fe/H]$=-0.078\pm0.003 \times R_\mathrm{GC} - 0.77\pm0.02$, while in
the outer regions it flattens to $\sim - 1.7$ dex. The presence of two
distinct components agrees with an ``inside-out'' galaxy formation
scenario such as for the closely related NGC 300 galaxy (Vlaji\'{c} et
al \cite{vla09}) and confirmed by Williams et al (\cite{wil09}). The
AGB gradient is steeper than that from HII regions supporting this
scenario.

The Magellanic Clouds show a different but linked metallicity history
influenced by their structure and dynamical interaction. It is easier
to interpret the metallicity of M33 that has, instead, evolved in
isolation.  The observation of the outer disc/halo population of the
Magellanic Clouds as traced here by AGB stars is not complete. Large
data bases can be exploited to search for AGBs $8-20$ kpc from the
centre. There are also fewer studies of chemical abundances and
kinematics in the SMC than in the LMC (van der Marel et al
\cite{vdm08}).

The upcoming VISTA survey of the Magellanic System (Cioni et al
\cite{vmc}) will provide new data to investigate the metallicity
evolution as well as the SFH, extinction and structure across the
system. The contemporary VISTA hemisphere survey will cover the
outermost regions of system. These near-infrared surveys will provide
targets for measuring abundances with current and future wide-field
spectrographs.

\begin{acknowledgements}
I deeply thank Marina Rejkuba for a critical reading of the
paper, and Sean Ryan, Ralf Napiwotzki and Janet Drew for interesting
scientific discussions contributing to its development.  
\end{acknowledgements}

\begin{appendix}
\section{Ca II triplet metallicity correction}
\label{app}

\begin{figure}
\label{metcal}
\resizebox{\hsize}{!}{\includegraphics{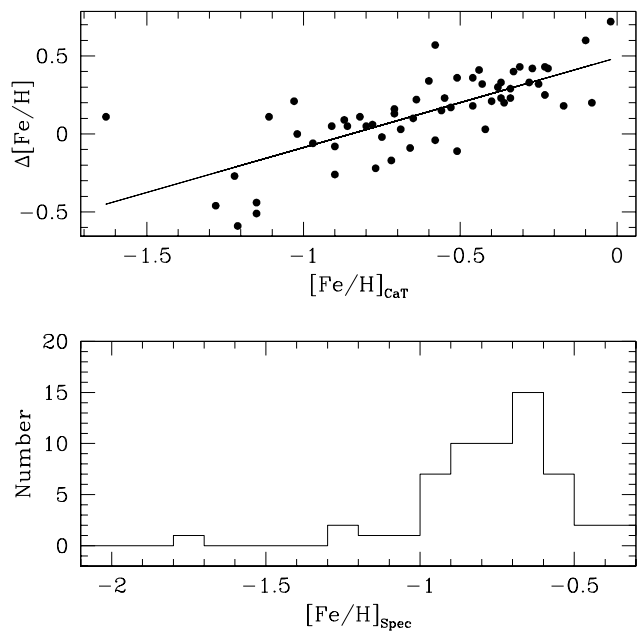}}
\caption{(Top) Iron abundance differences derived from the Ca II
  triplet and direct observation of iron lines as a function of Ca II
  triplet values (Pomp\'{e}ia et al \cite{pom08}). The line is the
  least square fit through the data points. (Bottom) Number
  distribution of iron abundances.}
\end{figure}

The study by Pomp\'{e}ia et al (\cite{pom08}) analyses the spectra of
$59$ field RGB stars deriving chemical abundances of iron, from both
FeI and FeII lines, and other $\alpha$ elements. The authors indicate
no systematics for [FeII/H] abundances but a systematics of $\sim0.1$
dex for [FeI/H] abundances.  Figure \ref{metcal} shows the comparison
between [Fe/H] values from iron lines and from the Ca II triplet
method, the data points are from their Tab.~2.  The least square fit
through these points corresponds to:
\begin{equation}
\Delta \mathrm{[Fe/H]}=-0.575\pm0.004\times \mathrm{[Fe/H]}_\mathrm{CaT}+0.48\pm0.01
\label{cor}
\end{equation}
where $\Delta$[Fe/H]$=$[Fe/H]$_\mathrm{CaT}-$[Fe/H]$_\mathrm{spec}$
and $\sigma_\mathrm{[Fe/H]}=0.15$ for each point. This equation, that
is well defined for [Fe/H]$<-1.3$ dex, has been used to correct [Fe/H]
values from the Ca II triplet, based on the same calibration (Carretta
\& Gratton \cite{cg97}).  The histogram of iron abundances peaks at a
disc stellar population with [Fe/H]$_\mathrm{spec}=-0.75\pm0.23$ dex.

Compared to Battaglia et al (\cite{bat08}), who studied RGB
metallicities in Sculptor and Fornax dwarf spheroidal galaxies,
Fig.~\ref{metcal} shows a strong gradient. These authors concluded
that Ca II triplet metallicities are overestimated by $\sim 0.1$ dex
at [Fe/H]$< -2.2$ dex, underestimated by $\sim 0.1-0.2$ dex at
[Fe/H]$> -1.2$ dex and have no trend for [Fe/H]$> -0.8$ dex in the
range $-2.5<$[Fe/H]$<-0.5$. Their RGB stars are $>8$ Gyr old
(Sculptor) and $3-6$ Gyr old (Fornax) while LMC RGB stars are mostly
young ($1-3$ Gyr old).  Here, Eq.~\ref{cor} is applied only to
  measurements for RGB stars younger than $\sim 5$ Gyr.  A different
SFH also implies a different Ca/Fe abundance, therefore, the Ca II
triplet is a good proxy for [Fe/H] whilst the age of RGB stars with
respect to the calibrating relation is appropriately considered (Pont
et al \cite{pon04}).

\section{C/M versus [Fe/H] calibration}
\label{cmfe}

Battinelli \& Demers (\cite{bat}) provide a calibration of the
metallicity of galaxies versus the C/M ratio. The major uncertainty in
their relation lies in the values of metallicity adopted from the
literature. These values refer to RGB stars, except for IC 10 where
[Fe/H] was converted from [O/H] in HII regions.

For most of the remaining galaxies the metallicity was estimated from
the colour of the RGB in $(V-I)_0$ using the relation by Da Costa \&
Armandroff (\cite{daa90}), or Lee (\cite{lee93}), defined at a
specific magnitudes below the TRGB. This method is sensitve to
metallicity but not to age, that does in fact represent the extent of
the RGB population. This section re-assesses the [Fe/H] versus
  C/M calibration using updated metallicity measurements for a more
  homogeneous population. Table \ref{bdtab} lists the C/M0+ values
  from Battinelli \& Demers (\cite{bat}) and the metallicities from
  different authors. The uncertainty in the metallicity values is the
  dispersion, $\sigma$, corresponding to half the width of the RGB at
  the calibrating colour. This quantity is a better indicator, than
  the statistical uncertainty, of the range of ages for the RGB
  population within each galaxy. 

\begin{figure}
\label{bdfig}
\resizebox{\hsize}{!}{\includegraphics{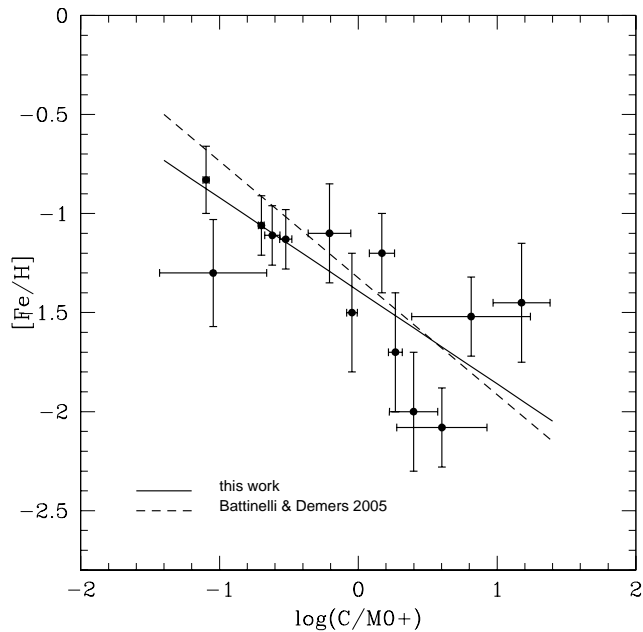}}
\caption{Metallicity as a function of C/M ratio. Points and their
  weighted least square fit line (solid line) are for RGB
  stars as discussed in the text. The dashed line is from 
  Battinelli \& Demers (\cite{bat}).}
\end{figure}

 The RGB parameters for the DDO 190 galaxy, well outside of the
  Local Group, were obtained by Battinelli \& Demers (\cite{bat06}),
  for the C/M ratio, and Aparicio \& Tikhonov (\cite{apt00}), for
  [Fe/H]. The [Fe/H] for NGC 3109 is the mean of the three values
  quoted in Battinelli \& Demers (\cite{bat}). 

In the case of NGC 6822 the RGB colour gives a metallicity $0.25$ dex
lower, Battinelli \& Demers (\cite{bat}) averaged this value with the
value obtained from the RGB slope in the near-IR domain.  For DDO 210
McConnachie et al (\cite{mc06b}) obtained [Fe/H]$=-1.3$ dex using the
RGB colour method and corresponding to an RGB population $4\pm2$ Gyr
old. This value is $0.6$ dex higher than the value used by Battinelli
\& Demers (\cite{bat}). A difference of $\sim 0.2$ dex is found for
NGC147, NGC 185 and NGC 205 by Butler \& Mart\'{i}nez-Delgado
(\cite{bum05}) who obtained a mean RGB metallicity of $-1.11$ dex and
$-1.06$, respectively.

For IC 1613 the metallicity was derived from the SFH resulting from
fitting colour-magnitude diagrams with synthetic diagrams produced
using stellar evolution models. These models provide metallicity in
terms of total metallicity, Z, that when converted to iron represents
an upper limit. The error in [Fe/H] given in Tab.~\ref{bdtab}
coincides with the FWHM of the RGB colour from Bernard et al
(\cite{ber07}).  A similar study has been done recently in M31
resulting in [Fe/H]$=-0.83$ dex (Brown et al \cite{br08}). This is a
mean value among fields $10-35$ kpc from the galaxy centre
corresponding to a RGB population with a mean age of $10.4\pm0.6$ Gyr.

Among the metallicities used by Battinelli \& Demers (\cite{bat}) that
of Leo I refers to spectroscopic iron lines observed in RGB stars.  In
the meantime there have been no other direct measures of iron lines,
in the current sample of Local Group galaxies, but several authors
have derived iron from the observation of the Ca II triplet in RGB
stars. These measurements are not included here because of two
uncertainties: the assumption on the [Ca/Fe] abundance ratio,
necessary to convert Ca II triplet abundances to iron, and the age of
the RGB stars (Sect.~\ref{app}).

Summarizing, updating metallicities obtained from the colour of the
RGB, including those of RGB populations derived from the SFH method
(IC 1613 only) and iron lines (Leo I only), and excluding IC 10 gives
the following relation:
\begin{equation}
\mathrm{[Fe/H]}=-1.39\pm0.06 -0.47\pm0.10 \times log(C/M0+)
\end{equation}
with $\chi ^2 = 1.1$.  Figure \ref{bdfig} shows the distribution of
the RGB points listed in Tab.~\ref{bdtab}, and their weighted least
square fit line as well as the line derived by Battinelli \& Demers
(\cite{bat}).  Weights account for both the error on the C/M0+ and
[Fe/H].

\begin{table}
\caption{RGB metallicity and C/M ratio}
\label{bdtab}
\[
\begin{array}{lcccc}
\hline
\noalign{\smallskip}
\mathrm{Galaxy} & \mathrm{[Fe/H]} & \sigma & \mathrm{C/M0+} & \mathrm{Reference} \\
 & \mathrm{(dex)} & \mathrm{(dex)} & & \\
\noalign{\smallskip}
\mathrm{DD0\,190}  & -2.00 & 0.30 & 2.5\pm1.0     & 1 \\
\mathrm{DDO\,210}  & -1.30 & 0.27 &  0.09\pm0.08  & 2 \\
\mathrm{IC\,1613}  & -1.20 & 0.20 &  1.48\pm0.31  & 3 \\
\mathrm{Leo\,I}    & -1.52 & 0.20 &   6.5\pm6.4   & 4 \\
\mathrm{M31}       & -0.83 & 0.17 & 0.080\pm0.004 & 5 \\
\mathrm{NGC\,147}  & -1.13 & 0.15 &  0.30\pm0.03  & 6 \\
\mathrm{NGC\,185}  & -1.11 & 0.15 &  0.24\pm0.03  & 6 \\
\mathrm{NGC\,205}  & -1.06 & 0.15 &  0.20\pm0.01  & 6 \\
\mathrm{NGC\,3109} & -1.70 & 0.30 &  1.85\pm0.21  & 7 \\
\mathrm{NGC\,6822} & -1.50 & 0.30 &  0.90\pm0.08  & 8 \\
\mathrm{Pegasus}   & -1.10 & 0.25 &  0.62\pm0.22  & 9 \\
\mathrm{SagDIG}    & -2.08 & 0.20 &   4.0\pm3.0   & 10 \\
\mathrm{WLM}       & -1.45 & 0.30 &  15.0\pm7.1   & 11 \\
\noalign{\smallskip}
\hline
\end{array}
\]

($1$) Aparicio \& Tikhonov \cite{apt00}; ($2$) McConnachie et al
\cite{mc06b}; ($3$) Skillman et al \cite{ski03}; ($4$) Tolstoy et al
\cite{tol03}; ($5$) Brown et al \cite{br08}; ($6$) Butler \&
Martinez-Delgado \cite{bum05}; ($7$) Battinelli \& Demers \cite{bat};
($8$) Gallart et al \cite{gal96}; ($9$) Aparicio \cite{apa94}; ($10$)
Momany et al \cite{mom02}; ($11$) Minniti \& Zijlstra \cite{min97}.
\end{table}
\end{appendix}


\begin{thebibliography}{}
\bibitem[1989]{and89}
  Anders, E., \& Grevesse, N., 1989, Geochim. Cosmochim. Acta, 53, 197
\bibitem[2000]{apt00}
  Aparicio, A., \& Tikhonov, N., 2000, AJ, 119, 2183
\bibitem[1994]{apa94}
  Aparicio, A., 1994, ApJ, 347, L27
\bibitem[1989]{arm99}
  Armandroff, T.E., 1989, AJ, 97, 375
\bibitem[2004]{asp04}
  Asplund, M., Grevesse, N., Suval, A.J., et al, 2004, A\&A, 417, 751
\bibitem[2007]{bar07}
  Barker, M. K., Sarajedini, A., Geisler, D., et al. 2007, AJ 133, 1125 
\bibitem[2008]{bat08}
  Battaglia, G., Irwin, M., Tolstoy, E., et al, 2008, MNRAS, 383, 183
\bibitem[2006]{bat06}
  Battinelli, P., \& Demers, S., 2006, A\&A, 447, 473
\bibitem[2005]{bat}
  Battinelli, P., \& Demers, S., 2005, A\&A, 434, 657
\bibitem[2009]{bat09}
  Battinelli, P., \& Demers, S., 2009, A\&A, 493, 1075
\bibitem[2004]{bek04}
  Bekki, K., Couch, W.J., Beasley, M.A., et al, 2004, ApJ, 610, L93
\bibitem[2007]{ber07}
  Bernard, E.J., Aparicio, A., Gallart, C., et al., 2007, AJ, 134, 1124
\bibitem[2007]{bes07}
  Besla, G., Kallivayalil, N., Hernquist, L., et al, 2007, ApJ, 668, 949
\bibitem[2006]{bor06}
  Borissova, J., Minniti, D., Rejkuba, m., \& Alves, D., 2006, A\&A,
  460, 459
\bibitem[2004]{bro04}
  Brooks, R.S., Wilson, C.D., \& Harris, W.E., 2004, AJ, 128, 237
\bibitem[2008]{br08}
  Brown, T.M., Beaton, M., Chiba, M., et al., 2008, ApJ, 685, L121
\bibitem[2005]{bum05}
  Butler, D.J., \& Mart\'{i}nez-Delgado, D., 2005, AJ, 129, 2217
\bibitem[2008a]{car08a}
  Carrera, R., Gallart, C., Hardy, E., et al, 2008a, AJ, 135, 836
\bibitem[2008]{car}
  Carrera, R., Gallart, C., Aparicio, A., et al. 2008, AJ, 136, 1039
\bibitem[1997]{cg97}
  Carretta, E., \& Gratton, R.G., 1997, A\&AS, 121, 95
\bibitem[2002]{cha02}
  Chandar, R., Bianchi, L., Ford, H.C., \& Sarajedini, A., 2002, AJ,
  564, 712
\bibitem[2001]{chi01}
  Chiappini, C., Matteucci, F., \& Romano, D., 2001, ApJ, 554, 1044
\bibitem[2000]{mor}
  Cioni, M.-R.L., Habing, H.J., \& Israel, F.P., 2000, A\&A, 358, L9
\bibitem[2000]{tip}
  Cioni, M.-R.L., van der Marel, R.P., Loup, C., \& Habing, H.J.,
  A\&A, 359, 601
\bibitem[2001]{cio01}
  Cioni, M.-R.L., Marquette, J.-B., Loup, C., et al, 2001, A\&A, 377, 945
\bibitem[2003]{cmr}
  Cioni, M.-R.L., \& Habing, H.J. 2003, A\&A, 402, 133
\bibitem[2003]{smcvar}
  Cioni, M.-R.L., Blommaert, J.A.D.L., Groenewegen, M.A.T., et al,
  2003, A\&A, 406, 51
\bibitem[2006a]{lfl}
  Cioni, M.-R.L., Girardi, L., Marigo, P., \& Habing, H.J. 2006a,
  A\&A, 448, 77 
\bibitem[2006b]{lfs}
  Cioni, M-R.L., Girardi, L., Marigo, P., \& Habing, H.J. 2006b, A\&A,
  452, 195
\bibitem[2008]{m33}
  Cioni, M.-R.L., Irwin, M., Ferguson, A.M.N., et al. 2008, A\&A 487,
  131 
\bibitem[2009]{vmc}
  Cioni, M.-R.L., Bekki, K., Clementini, G., et al, 2008,
  PASA, 25, 121
\bibitem[2000]{col00}
  Cole, A.A., Smecker-Hane, T.A., \& Gallagher, J.S., 2000, AJ, 120, 1808
\bibitem[2005]{col05}
  Cole, A.A., Tolstoy, E., Gallagher, J.S., \& Smecker-Hane, T.A.,
  2005, AJ, 129, 1465
\bibitem[1990]{daa90}
  Da Costa, G.S., \& Armandroff, T.E., 1990, AJ, 100, 162
\bibitem[1998]{dac98}
  Da Costa, G.S., \& Hatzidimitriou, D., 1998, AJ, 115, 1934
\bibitem[2003]{dav03}
  Davidge, T.J., 2003, AJ, 125, 3046
\bibitem[1984]{duf84}
  Dufour, R.J., 1984, Structure and evolution of the MC's, ed.~S.~van
  den Bergh, \& K.S.~de Boer, IAU Symp.~108, 353
\bibitem[2008]{duf08}
  Dufton, P.L., Ryans, R.S.I., Thompson, H.M.A., \& Street, R.A.,
  2008, MNRAS, 385, 2261
\bibitem[1993]{edv93}
  Edvardsson, B., Andersen, J., Gustafsson, B., et al., 1993, A\&A,
  275, 101
\bibitem[2007]{fer07}
  Ferguson, A., Irwin, M., Chapman, S., et al, 2007, Island Universes,
  ASSP Springer, 239 
\bibitem[1995]{fri95}
  Friedli, D., \& Benz, W., 1995, A\&A, 301, 649
\bibitem[1996]{gal96}
  Gallart, C., Aparicio, A., \& Vilchez, J.M., 1996, AJ, 112, 1928
\bibitem[2008]{gal08}
  Gallart, C., Stetson, P.B., Meschin, I., et al, 2008, ApJ, 682, L89
\bibitem[2003]{gla03}
  Glass, I.S., \& Schultheis, M., 2003, MNRAS, 345, 39
\bibitem[2009]{god09}
  Gonidakis, I., Livanou, E., Kontizas, E., et al, 2009, A\&A, {\it
  accepted}, astro-ph/0812.0880
\bibitem[2009]{gor09}
  Gordon, K.D., Bot, C., Muller, E., et al, 2009, ApJ, 690, L76 
\bibitem[2006]{gro06}
  Grocholski, A.J., Cole, A.A., Sarajedini, A., et al, 2006, AJ, 132, 1630
\bibitem[2007]{gro07}
  Grocholski, A.J., Sarajedini, A., Olsen, K., et al, 2007, AJ, 134, 680
\bibitem[2004]{har04}
  Harris, J., \& Zaritsky, D., 2004, AJ, 127, 1531
\bibitem[2007]{har07}
  Harris, J., 2007, ApJ, 658, 345
\bibitem[2000]{hil00}
  Hill, V., Fran\c{c}ois, P., Spite, M., et al 2000, A\&A, 364, L19
\bibitem[1991]{hug91}
  Hughes, S>M.G., Wood, P.R., \& Reid, N., 1991, AJ, 101, 1304
\bibitem[2007]{idi07}
  Idiart, T.P., Maciel, W.J., \& Costa, R.D.D., 2007, A\&A, 472, 101
\bibitem[2002]{jac02}
  Jacoby, J.H., \& De Marco, O., 2002, AJ, 123, 269
\bibitem[2006b]{kal06b}
  Kallivayalil, N., van der Marel, R.P., Alcock, C., et al, 2006b,
  ApJ, 638, 772
\bibitem[2006a]{kal06a}
  Kallivayalil, N., van der Marel, R.P., \& Alcock, C., 2006a, ApJ,
  652, 121
\bibitem[1999]{kar99}
  Karachentsev, I., Aparicio, A., \& Makarova, L., 1999, A\&A, 352, 363
\bibitem[2002]{kim02}
  Kim, M., Kim, E., Lee, M.G., et al, 2002, AJ, 123, 244
\bibitem[2000]{king}
  King, J.R., 2000, AJ, 120, 1056
\bibitem[1993]{lee93}
  Lee, M.G., 1993, ApJ, 408, 409
\bibitem[2008]{leh08}
  Lehner, R., Howk, J.C., Keenan, F.P., \& J.V., Smoker, 2008, ApJ,
  678, 219
\bibitem[2006]{lei06}
  Leisy, P., \& Dennefled, M., 2006, A\&A, 456, 451
\bibitem[2004]{ma}
  Ma, J., Zhou, X., \& Chen, J., 2004, A\&A, 413, 563
\bibitem[2007]{mag07}
  Magrini, L., Corbelli, E., \& Galli, D., A\&A, 470, 843
\bibitem[2009]{mag09}
  Magrini, L., Stanghellini, L., \& Villaver, E. 2009, ApJ, {\it
  submitted}, astro-ph/0901.2273
\bibitem[1999]{ma99}
  Marigo, P., Girardi, L., \& Bressan, A., 1999, A\&A, 344, 123
\bibitem[2005]{maj05}
  Majewski, S.M., Frinchaboy, P.M., Peter, M., et al., 2005, AJ, 130, 2677
\bibitem[1994]{roy94}
  Martin, P., \& Roy, J.-R., 1994, ApJ, 424, 599
\bibitem[2008]{mas08}
  Mastropietro, C., 2008, The Magellanic System: Stars, Gas, and
  Galaxies, ed.~J.Th. van Loon \& J.M. Oliveira, IAU Symp.~256, {\it in press}
\bibitem[2006b]{mc06b}
  McConnachie, A.W., Alan, W., Arimoto, N., et al., 2006b, MNRAS, 373, 715
\bibitem[2006a]{mc06a}
  McConnachie, A.W., Alan W., Chapman, S.C., et al, 2006a, ApJ, 647, L25 
\bibitem[1996]{mcl96}
  McLean, I.S., \& Liu, T., 1996, ApJ, 456, 499
\bibitem[1997]{min97}
  Minniti, D., Zijlstra, A.A., 1997, AJ, 114, 147
\bibitem[2003]{min03}
  Minniti, D., Borissova, J., Rejkuba, M., et al, 2003, Sci., 301, 1508
\bibitem[2002]{mom02}
  Momany, Y.,Held, E.V., Saviane, I., \& Rizzi, L., 2002, A\&A, 384, 393
\bibitem[2008]{nid08}
  Nidever, D.L., Majewski, S.R., \& Butler, W.B., 2008, AJ, 679, 432
\bibitem[2007]{noe07}
  N\"{o}el, N.E., \& Gallart, C., 2007, AJ, 665, L23
\bibitem[2009]{par09}
  Parisi, M.C., Grocholski, A.J., Geisler, D., et al, 2009, AJ, {\it
  submitted}, astro-ph/0808.0018
\bibitem[2003]{pat}
  Paturel, G., Petit, C., Prugniel, P., et al. 2003, A\&A, 412, 45 
\bibitem[2005]{pia05}
  Piatti, A.E., Sarajedini, A., Geisler, D., et al, 2005, MNRAS, 358, 1215
\bibitem[2008]{pom08}
  Pomp\'{e}ia, L., Hill, V., Spite, M., et al., 2008, A\&A, 480, 379
\bibitem[2004]{pon04}
  Pont, F., Zinn, R., Gallart, C., et al, 2004, AJ, 127, 840
\bibitem[2006]{pne}
  Reid,W.A., \& Parker, Q.A., 2006, MNRAS, 373, 521
\bibitem[2008]{ros08}
  Ro\v{s}kar, R., Debattista, V.P., Stinson, G.S., et al, 2008, ApJ,
  675, L65
\bibitem[2008]{rosi08}
  Rosolowsky, E., \& Simon, J.D., 2008, ApJ, 675, 1213
\bibitem[2000]{sar00}
  Sarajedini, A., Geisler, D., Schommer, R., \& Harding, P., 2000, AJ,
  120, 2437
\bibitem[2006]{sar06}
  Sarajedini, A., Barker, M.K., Geisler, D., et al, 2006, AJ, 132, 1361
\bibitem[1991]{scho91}
  Schommer, R.A., Christian, C.A., Caldwell, N., et al, 1991, AJ, 101, 873
\bibitem[2002]{sel02}
  Sellwood, J.A., \& Binney, J.J., 2002, MNRAS, 336, 785
\bibitem[2003]{ski03}
  Skillman, E.D., Tolstoy, E., Cole, A.A., et al., 2003, ApJ, 596, 253
\bibitem[2005]{sou05}
  Soubiran, C., \& Girard, P., 2005, A\&A, 438, 139
\bibitem[2006]{sub06}
  Subramaniam, A., 2006, A\&A, 449, 101
\bibitem[2009]{sub09}
  Subramanian, S., \& Subramaniam, A., 2009, A\&A, {\it accepted}, astro-ph/0809.4362
\bibitem[2004]{tie04}
  Tiede, G.P., Sarajedini, A., \& Barker, M.K., 2004, AJ, 128, 224 
\bibitem[2003]{tol03}
  Tolstoy, E., Venn, K.A, Shetrone, M., et al., 2003, AJ, 125, 707
\bibitem[2008]{tos08}
  Tosi, M., Gallagher, J., \& Sabbi, E., et al, 2008, Low-Metallicity
  Star Formation: From the First Stars to Dwarf Galaxies,
  ed.~L.K. Hunt, S. Madden \& R. Schneider, IAU Symp.~255, 381
\bibitem[2001]{vdm}
  van der Marel, R.P., \& Cioni, M.-R.L., 2001, AJ 122, 1807
\bibitem[2001]{vdm1}
  van der Marel, R.P., 2001, AJ, 122, 1827
\bibitem[2002]{vdm02}
  van der Marel, R.P., Alves, D.R., Hardy, E., \& Suntzeff, N.B.,
  2002, AJ, 124, 2639
\bibitem[2008]{vdm08}
  van der Marel, R.P., Kallivayalil, N., \& Besla, G., 2008, The
  Magellanic System: Stars, Gas, and Galaxies, ed.~J.Th. van Loon \&
  J.M. Oliveira, IAU Symp.~256, {\it in press}
\bibitem[2009]{vla09}
  Vlaji\'{c}, M., Bland-Hawthorn, J., \& Freeman, K.C., 2009, ApJ,
  {\it accepted}
\bibitem[2001]{wak01}
  Wakker, B.P., 2001, ApJS, 136, 463
\bibitem[2009]{wil09} 
  Williams, B.F., Dalcanton J.J., Dolphin, A.E.,
  et al, 2009, ApJ, {\it accepted}, astro-ph/0902.3460
\bibitem[1994]{zar94}
  Zaritsky, D., Kennicutt, R.C., Jr., \& Hucra, J.P., 1994, ApJ, 420, 87
\bibitem[2002]{zar02}
  Zaritsky, D., Harris, J., Thompson, I.B., et al, 2002, AJ, 123, 855
\bibitem[2004]{zar04}
  Zaritsky, D., Harris, J., Thompson, I.B., \& Grebel, E.K., 2004, AJ,
  128, 1606  

\end{thebibliography}
\end{document}